\newcommand{\CLASSINPUTtoptextmargin}{19mm}%
\newcommand{\CLASSINPUTbottomtextmargin}{43mm}%
\newcommand{\CLASSINPUTinnersidemargin}{12.9mm}%
\newcommand{\CLASSINPUToutersidemargin}{12.9mm}%
\documentclass[conference,10pt,a4paper]{IEEEtran}%
\pdfoutput=1
\usepackage{amsmath}%
\usepackage{cite}
\usepackage{times}%
\usepackage{graphicx}%
\usepackage{orcidlink}
\usepackage{url}
\usepackage{multirow}%
\usepackage{float}%
\usepackage{subfig}%
\usepackage{iftex}%
\usepackage{amssymb,amsfonts,yhmath,mathrsfs}
\usepackage{bm}

\ifPDFTeX%
\usepackage{t1enc}%
\usepackage{times}%
\fi%
\ifLuaTeX%
\usepackage{fontspec}%
\fi%
\makeatletter

\def\@maketitle{\newpage
\bgroup\par\addvspace{0.5\baselineskip}\centering%
\ifCLASSOPTIONtechnote
   {\bfseries\large\@IEEEcompsoconly{\sffamily}\@title\par}\vskip 1.3em{\lineskip .5em\@IEEEcompsoconly{\sffamily}\@author
   \@IEEEspecialpapernotice\par{\@IEEEcompsoconly{\vskip 1.5em\relax
   \@IEEEtitleabstractindextextbox{\@IEEEtitleabstractindextext}\par
   \hfill\@IEEEcompsocdiamondline\hfill\hbox{}\par}}}\relax
\else
   \vskip0.2em{\EuMWtitlesize\ifCLASSOPTIONtransmag\bfseries\LARGE\fi\@IEEEcompsoconly{\sffamily}\@IEEEcompsocconfonly{\normalfont\normalsize\vskip 2\@IEEEnormalsizeunitybaselineskip
   \bfseries\Large}\@title\par}\vskip1.0em\par
   \ifCLASSOPTIONconference%
      {\@IEEEspecialpapernotice\mbox{}\vskip\@IEEEauthorblockconfadjspace%
       \mbox{}\hfill\begin{@IEEEauthorhalign}\@author\end{@IEEEauthorhalign}\hfill\mbox{}\par}\relax
   \else
      \ifCLASSOPTIONpeerreviewca
         {\@IEEEcompsoconly{\sffamily}\@IEEEspecialpapernotice\mbox{}\vskip\@IEEEauthorblockconfadjspace%
          \mbox{}\hfill\begin{@IEEEauthorhalign}\@author\end{@IEEEauthorhalign}\hfill\mbox{}\par
          {\@IEEEcompsoconly{\vskip 1.5em\relax
           \@IEEEtitleabstractindextextbox{\@IEEEtitleabstractindextext}\par\hfill
           \@IEEEcompsocdiamondline\hfill\hbox{}\par}}}\relax
      \else
         \ifCLASSOPTIONtransmag
           {\@IEEEspecialpapernotice\mbox{}\vskip\@IEEEauthorblockconfadjspace%
            \mbox{}\hfill\begin{@IEEEauthorhalign}\@author\end{@IEEEauthorhalign}\hfill\mbox{}\par
           {\vspace{0.5\baselineskip}\relax\@IEEEtitleabstractindextextbox{\@IEEEtitleabstractindextext}\vspace{-1\baselineskip}\par}}\relax
         \else
           {\lineskip.5em\@IEEEcompsoconly{\sffamily}\sublargesize\@author\@IEEEspecialpapernotice\par
           {\@IEEEcompsoconly{\vskip 1.5em\relax
            \@IEEEtitleabstractindextextbox{\@IEEEtitleabstractindextext}\par\hfill
            \@IEEEcompsocdiamondline\hfill\hbox{}\par}}}\relax
         \fi
      \fi
   \fi
\fi\par\addvspace{0.0\baselineskip}\egroup}

\def\EuMWtitlesize{\@setfontsize{\EuMWtitlesize}{24}{24pt}}
\def\EuMWauthorsize{\@setfontsize{\EuMWauthorsize}{11}{11pt}}
\def\EuMWaffilsize{\@setfontsize{\EuMWaffilsize}{10}{10pt}}
\def\EuMWcaptionsize{\@setfontsize{\EuMWcaptionsize}{9}{10pt}}
\def\EuMWbibsize{\@setfontsize{\EuMWbibsize}{8}{10pt}}

\def\@IEEEauthorblockNstyle{\EuMWauthorsize\@IEEEcompsocnotconfonly{\sffamily}\@IEEEcompsocconfonly{\large}}
\def\@IEEEauthorblockAstyle{\EuMWaffilsize\@IEEEcompsocnotconfonly{\sffamily}\@IEEEcompsocconfonly{\itshape}\@IEEEcompsocconfonly{\large}}
\def\@IEEEauthordefaulttextstyle{\EuMWauthorsize\@IEEEcompsocnotconfonly{\sffamily}\sublargesize}

\def\thebibliography#1{\section*{\refname}%
    \addcontentsline{toc}{section}{\refname}%
    \EuMWbibsize\@IEEEcompsocconfonly{\small}\vskip 0.3\baselineskip plus 0.1\baselineskip minus 0.1\baselineskip
    \list{\@biblabel{\@arabic\c@enumiv}}%
    {\settowidth\labelwidth{\@biblabel{#1}}%
    \leftmargin\labelwidth
    \advance\leftmargin\labelsep\relax
    \itemsep \IEEEbibitemsep\relax
    \usecounter{enumiv}%
    \let\p@enumiv\@empty
    \renewcommand\theenumiv{\@arabic\c@enumiv}}%
    \let\@IEEElatexbibitem\bibitem%
    \def\bibitem{\@IEEEbibitemprefix\@IEEElatexbibitem}%
\def\newblock{\hskip .11em plus .33em minus .07em}%
\ifCLASSOPTIONtechnote\sloppy\clubpenalty4000\widowpenalty4000\interlinepenalty100%
\else\sloppy\clubpenalty4000\widowpenalty4000\interlinepenalty500\fi%
    \sfcode`\.=1000\relax}

\long\def\@makecaption#1#2{%
\ifx\@captype\@IEEEtablestring%
\par\@IEEEtabletopskipstrut
\else
\@IEEEfigurecaptionsepspace
\fi
\setbox\@tempboxa\hbox{\normalfont\footnotesize {#1.}\nobreakspace\nobreakspace #2}%
\ifdim \wd\@tempboxa >\hsize%
\setbox\@tempboxa\hbox{\normalfont\footnotesize {#1.}\nobreakspace\nobreakspace}%
\parbox[t]{\hsize}{\normalfont\footnotesize\noindent\unhbox\@tempboxa#2}%
\else
\ifCLASSOPTIONconference \hbox to\hsize{\normalfont\footnotesize\hfil\box\@tempboxa\hfil}%
\else \hbox to\hsize{\normalfont\footnotesize\box\@tempboxa\hfil}%
\fi\fi
\ifx\@captype\@IEEEtablestring%
\@IEEEtablecaptionsepspace
\else
\fi}

\newlength\tablecaptiontotableskip
\newlength\figuretocaptionskip
\def\@IEEEfigurecaptionsepspace{\vskip\figuretocaptionskip\relax}%
\def\@IEEEtablecaptionsepspace{\vskip\tablecaptiontotableskip\relax}%

\def\abstract{\normalfont%
\@IEEEabskeysecsize\bfseries\textit{\abstractname}\,\bfseries\textit{---}\,%
\@IEEEgobbleleadPARNLSP}%

\def\IEEEkeywords{\normalfont%
\@IEEEabskeysecsize\bfseries\textit{\IEEEkeywordsname}\,\bfseries\textit{---}\,%
\@IEEEgobbleleadPARNLSP}%
\def\endIEEEkeywords{\relax\vspace{0.67ex}%
\par\if@twocolumn\else\endquotation\fi%
\normalsize\normalfont}%

\DeclareRobustCommand*{\EuMWauthorrefmark}[1]{\raisebox{0pt}[0pt][0pt]{\textsuperscript{#1}}}%
\def\@IEEEauthorblockNtopspace{0ex}
\def\@IEEEauthorblockAtopspace{1mm}
\def\tablename{Table}
\def\thetable{\arabic{table}}
\def\IEEEkeywordsname{Keywords}
\def\subsubsection{\@startsection{subsubsection}{3}{\z@}{1.5ex plus 1.5ex minus 0.5ex}%
{0.7ex plus .5ex minus 0ex}{\normalfont\normalsize\itshape}}%
\newlength{\CPheadmatchindent}%
\def\@seccntformat#1{\hbox to\CPheadmatchindent{\csname the#1dis\endcsname}\hskip 0.1em \relax}
\IEEEilabelindentA \parindent
\IEEEilabelindent \IEEEilabelindentA
\IEEEelabelindent \parindent
\IEEEdlabelindent \parindent
\IEEElabelindent \parindent
\makeatother

\usepackage{siunitx}

\usepackage{hyperref}

\begin{document}
\raggedbottom
\title{Bistatic Micro-Doppler Analysis of \\a Vertical Takeoff and Landing (VTOL) Drone \\in ICAS Framework}
\author{\IEEEauthorblockN{
Heraldo Cesar Alves Costa\EuMWauthorrefmark{*}\orcidlink{0009-0002-6186-5780},
Saw James Myint\EuMWauthorrefmark{*}\orcidlink{0009-0007-3788-7126},
Carsten Andrich\EuMWauthorrefmark{*}\orcidlink{0000-0002-4795-3517},
Sebastian W. Giehl\EuMWauthorrefmark{*}\orcidlink{0009-0008-1672-1351}, \\
Dieter Novotny \EuMWauthorrefmark{$\dagger$}, 
Julia Beuster\EuMWauthorrefmark{*}\orcidlink{0000-0003-1887-4278},
Christian Schneider\EuMWauthorrefmark{*}\orcidlink{0000-0003-1833-4562}, %
Reiner S. Thom\"a\EuMWauthorrefmark{*}\orcidlink{0000-0002-9254-814X}  
}   

\IEEEauthorblockA{
\EuMWauthorrefmark{*}Institute for Information Technology and Thuringian Center of Innovation in Mobility,\\Technische Universität Ilmenau, Ilmenau, Germany\\
\EuMWauthorrefmark{$\dagger$}AeroDCS Gmbh, Koblenz, Germany\\
heraldo-cesar.alves-costa@tu-ilmenau.de
}
}

\maketitle
\begin{abstract}
Integrated Communication and Sensing (ICAS) is a key technology that enables sensing functionalities within the next-generation mobile communication (6G).
Joint design and optimization of both functionalities could allow coexistence, therefore it advances toward joint signal processing and using the same hardware platform and common spectrum.
Contributing to ICAS sensing, this paper presents the measurement and analysis of the micro-Doppler signature of Vertical Takeoff and Landing (VTOL) drones.
Measurement is performed with an OFDM-like communication signal and bistatic constellation, which is a typical case in ICAS scenarios. 
This work shows that micro-Doppler signatures can be used to precisely distinguish flight modes, such as take-off, landing, hovering, transition, and cruising.
\end{abstract}
\begin{IEEEkeywords}
bistatic, micro-Doppler, OFDM, ICAS, flight modes, VTOL.
\end{IEEEkeywords}

\section{Introduction}\label{sec:Introduction}
Identification of different flight modes is important for detection, tracking, countermeasures, and air traffic management.
Flight modes can generally be categorized as take-off, landing, transition, and cruise.
\begin{figure}[b]
\centering
    \includegraphics[width=0.48\textwidth]{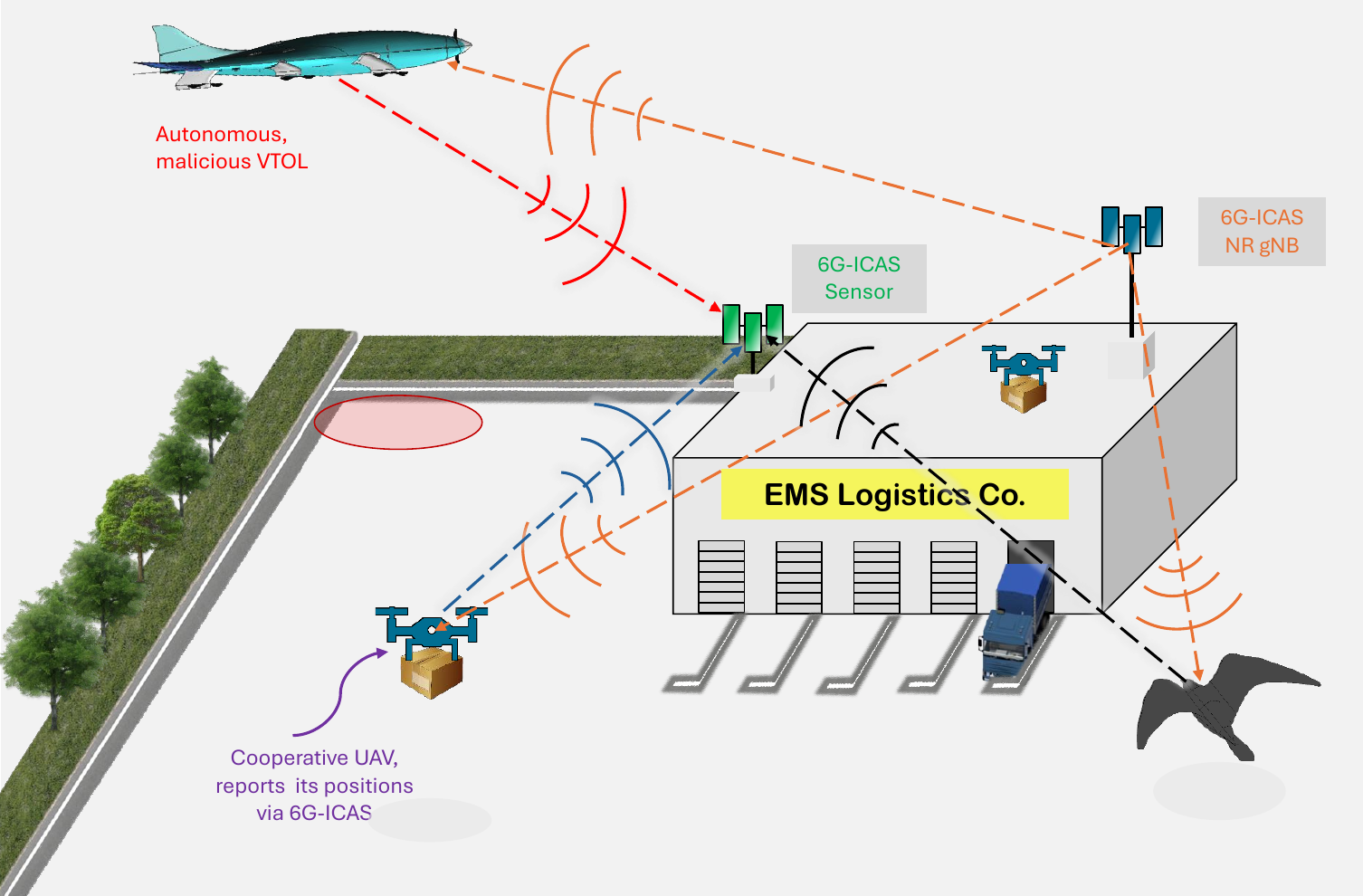}
    \caption{Example scenario of the ICAS-based surveillance system}
    \label{fig_Scenario}
\end{figure}

In \cite{Jakub2023} and \cite{Perrichon2024}, identification is done by methods based on machine learning or statistics.
However, they used flight data provided by onboard sensors.
In \cite{Strong2021}, they classified the flight modes with Hidden Markov Model using position, velocity, and acceleration from radar measurement.
Radar Cross Section (RCS) and micro-Doppler, in \cite{Gong2022} and \cite{Yan2023}, are used to classify a VTOL, a Quad-Rotor drone, and a fix-wing drone. 
The bistatic micro-Doppler, in \cite{Hoffmann2016}, is utilized to discriminate drones from the clutter, and therefore enhance detection probability.

A new possibility to acquire data for flight mode classification is the sensing technology of Integrated Communication and Sensing (ICAS).
ICAS is a technology concept that integrates sensing functionalities into the next generation of wireless communication (6G), which are co-designed for mutual benefits, i.e., via communication-assisted sensing and sensing-assisted communication \cite{Liu2022}.

An example scenario of an ICAS-based surveillance system is illustrated in \autoref{fig_Scenario}.
The cooperative drones periodically report their flight information to the base station.
When uncooperative drones enter the surveillance area, they need to be identified.
In this case, the sensing data can be used to perform activities, such as detection, tracking, target classification, and flight mode classification.

This paper focuses on the classification of the flight modes: (a) take-off/landing (b) transition, and (c) cruise of a VTOL drone using its micro-Doppler signature.
We consider bistatic constellation and the transmit signal similar to orthogonal frequency-division multiplexing (OFDM) under ICAS framework.

In section II, the airframe of a VTOL, its propeller composition, and basic flight modes are presented.
The bistatic micro-Doppler measurement and detail analysis can be found in Section III.

\section{VTOL Drone}\label{sec:VTOL}
\begin{figure}[t]
\centering
    \includegraphics[width=0.45\textwidth, height=1.8in]{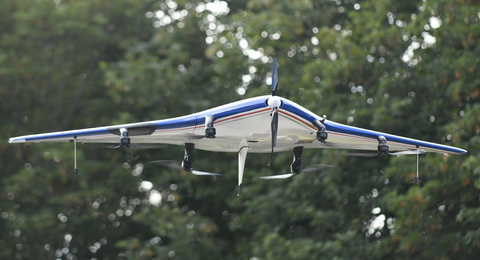}
    \caption{Vertical Take-Off and Landing (VTOL) drone.}
    \label{fig_VTOL}
\end{figure}

\begin{figure}[t]
\centering
    \includegraphics[width=0.45\textwidth, height=1.8in]{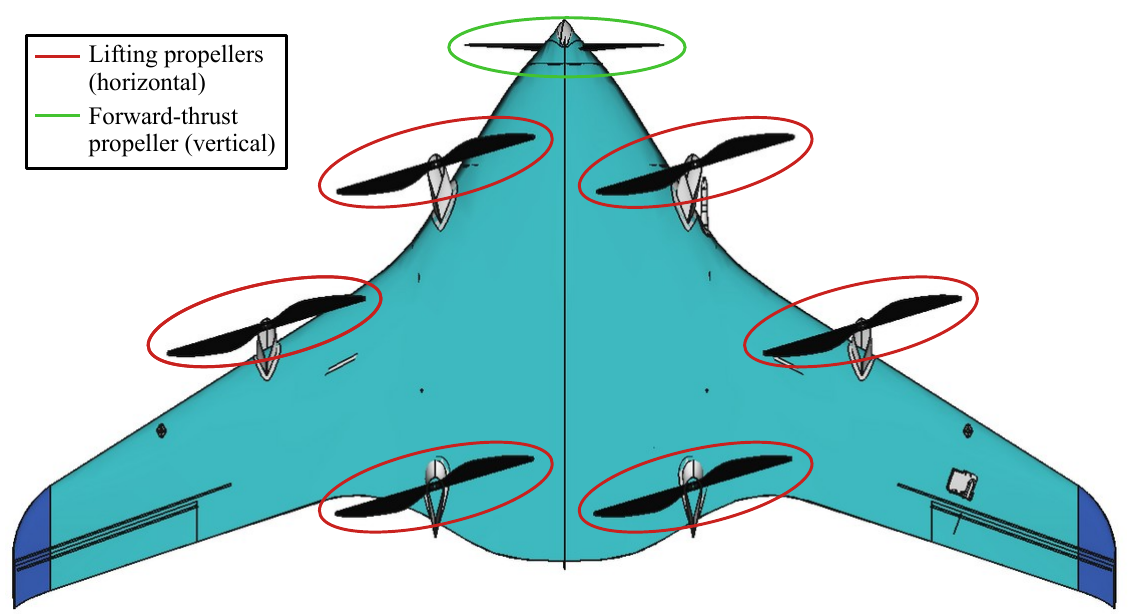} %
    \caption{VTOL propellers.}
    \label{fig_VTOL_propellers}
\end{figure}

The fixed-wing VTOL depicted in \autoref{fig_VTOL} is a versatile drone equipped with a total of seven propellers, which are six lifting (horizontal) propellers and one forward-thrust (vertical) propeller, as shown in \autoref{fig_VTOL_propellers}.
This design is a typical VTOL that combines multi-rotor and fixed-wing characteristics, using separate sets of propellers for different phases of flight.

\begin{figure}[b]
\centering
    \includegraphics[width=0.48\textwidth, height=1.6in]{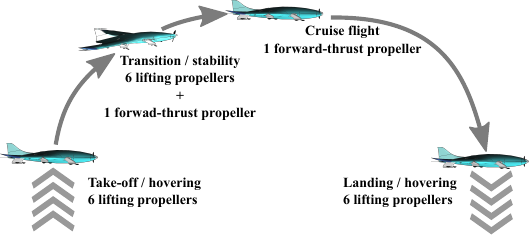}
    \caption{VTOL flight modes.}
    \label{fig_VTOL_flight_modes}
\end{figure}

During vertical flight modes such as take-off, hovering, and landing, the six lifting propellers provide the necessary vertical lift \cite{cakici2016design}.
In this mode, the forward-thrust propeller can be temporarily activated to improve stability.
When this VTOL transitions to forward flight, the six lifting propellers and the forward-thrust propeller operate simultaneously.
During the cruise flight, the forward-thrust propeller becomes the primary means of propulsion and the six lifting propellers are usually turned off.
In this mode, the aerodynamic design of the wings generates lift.
\autoref{fig_VTOL_flight_modes} shows all these flight modes.

The use of different propellers for different flight modes allows to classify the flight mode by checking which propellers are active.
This information can be obtained from the micro-Doppler signature, as will be discussed in \autoref{sec:Micro-Doppler}.

\section{Micro-Doppler Signature}\label{sec:Micro-Doppler}
Micro-Doppler refers to the modulation on the radar returns due to Doppler frequency shifts caused by the local movements of the target, such as rotations or vibrations.
Therefore, micro-Doppler provides valuable information about the target's structure and motion, which can be used to identify which kind of internal movements a target presents.
This information can be used for many applications, such as identify the type of target, distinguish it from other objects, or differentiate types of parts movements.

The measurements in this work were tailored to provide conditions present in ICAS frameworks including bistatic geometry and OFDM-like waveform \cite{Thoma2}.

\subsection{VTOL Bistatic Micro-Doppler Measurements} \label{ssec:MicroDoppler_measurements}
\begin{figure}[b]
    \centering
    \includegraphics[width=0.45\textwidth, height=1.7in]{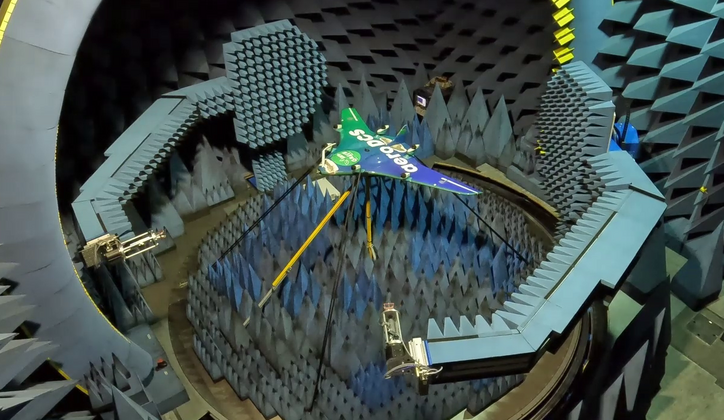}
    \caption{Micro-Doppler measurement in BIRA.}
    \label{fig_BiRa}
\end{figure}

For this analysis, multiple bistatic micro-Doppler measurements with different aspect angles and employing OFDM-like waveform were performed.
A solution for this kind of measurement is BIRA measurement system \cite{andrich2024bira}, illustrated in Fig. \ref{fig_BiRa}. 
BIRA has a wideband channel sounder, which can be used associated to the software-defined radio (SDR) system architecture described in \cite{rfsoc_paper}, allowing high range resolution (HRR) micro-Doppler measurements, with instantaneous bandwidth broader than \qty{2}{\GHz}.

{%
\setlength{\tabcolsep}{2mm}%
\renewcommand{\arraystretch}{1.2}%
\newcommand{\CPcolumnonewidth}{not used}%
\newcommand{\CPcolumntwowidth}{21mm}%
\newcommand{\CPcolumnthreewidth}{12mm}%
\newcommand{\CPcolumnfourwidth}{33mm}%
\begin{table}[t]
    \caption{Micro-Doppler Measurement Setup.}
    \small%
    \centering
    \begin{tabular}{|l|l|l|}\hline
        \multirow{4}{28mm}{{\bfseries Measurement System}}  & Waveform & OFDM-like\\ 
            \cline{2-3}
            & Central frequency & \qty{7}{\GHz} \\
            \cline{2-3}
            & Bandwidth & \qty{2.4}{\GHz} \\
            \cline{2-3}
            & Polarization & HH \\
            \hline
        \multirow{3}{28mm}{\parbox{24mm}{{\bfseries OFDM Settings}}} & Total number of carriers & 2500 \\
            \cline{2-3}
            & Carriers with energy & 2048 \\
            \cline{2-3}
            & Symbol duration & $ \approx \qty{1}{\us}$ \\
            \hline
        \multirow{5}{28mm}{\parbox{24mm}{{\bfseries VTOL Drone}}} & Number of propellers & 7 \\
            \cline{2-3}
            & Blades per propeller & 2 \\
            \cline{2-3}
            & Propeller radius & \qty{28.19}{\cm}\\
            \cline{2-3}
            & Propeller material & carbon fiber \\
        \hline
    \end{tabular}
    \label{tab_MicroDoppler_measurement_setup}
\end{table}
}

All measurements were made with one antenna pointing to the forward-thrust propeller, as shown in \autoref{fig_BiRa}, and with other antenna in different positions, forming many combinations of bistatic angles.
\autoref{tab_MicroDoppler_measurement_setup} details the measurement setup for this work.
A wideband OFDM-based transmit signal, called Newman sequence with constant spectral magnitude and minimal crest-factor \cite{Boyd1986}, was used.

Consider a system that transmits M symbols, with N samples (frequency bins) each.
For micro-Doppler analysis, the received signal was processed in slow-time, i.e., over different symbols. %
To achieve this, for each symbol $m$ $(0 \leq m < M)$, first, both the received signal $y(\tau, t)$ and a copy of the transmitted symbol $x(\tau, t)$, both in the time domain, are converted into the frequency domain ($X(f)$ and $Y(f)$), respectively), where $t = mT$ is the slow time, $T$ is the period between consecutive symbols, $\tau$ is the delay (travel time Tx-target-Rx), and $f = n \frac{f_s}{N}$ $(0 \leq n < N)$ are the baseband frequencies, where $f_s$ stands for the sampling frequency.
Then, the pilot frequencies are excluded, and the channel is estimated by doing $H(f) = \frac{Y(f)}{X(f)}$\cite{braun2014ofdm}.

After channel estimation, target detection is performed. Then, the returns from the target are arranged into a vector over different symbols, forming a slow-time profile, used to calculate Spectrogram, as detailed in \cite{costa2024bistatic}, and to plot the range-Doppler map.

\begin{figure}[b]
    \centering
    \includegraphics[width=0.45\textwidth, height=2in]{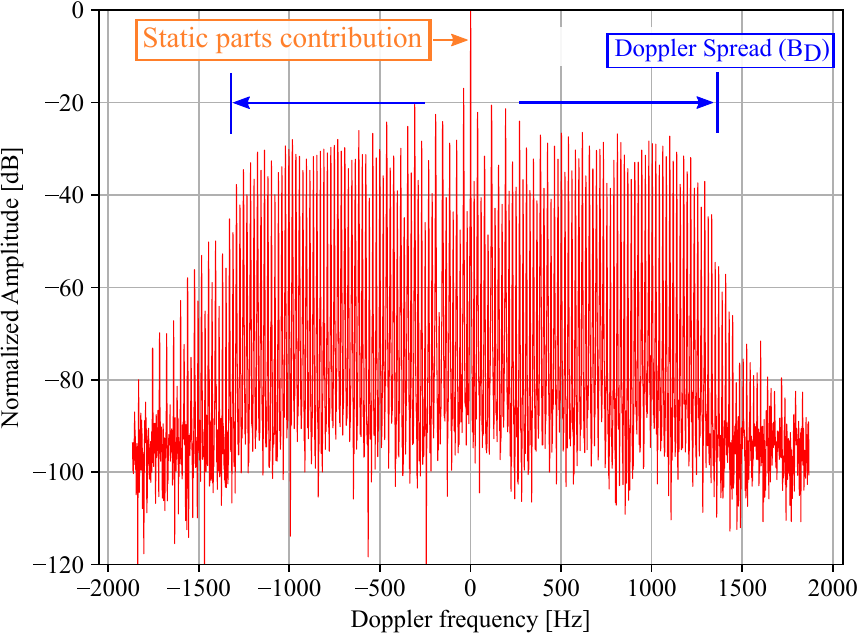}
    \caption{Frequency domain micro-Doppler signature of a single VTOL rotating propeller (16384 symbols, bistatic angle = $\ang{60}$).}
    \label{fig_FFT_propeller}
\end{figure}

\subsection{Single Propeller Micro-Doppler} \label{ssec:single_propeller}
A micro-Doppler signature of one horizontal rotating propeller in the frequency domain is presented in \autoref{fig_FFT_propeller}. 
As explained in \cite{costa2024bistatic}, the micro-Doppler signature of a single propeller in frequency domain presents a strong peak centered in $f = \qty{0}{\Hz}$ and many spikes on both sides of this central peak can be seen above the noisy basis. 
The central peak occurs due to the static parts of the drone, where $f = \qty{0}{\Hz}$ is the Doppler frequency of these static parts, while the other spikes are caused by the rotating propeller, and are separated by $\Delta f = N_b f_{rot}$, where $f_{rot}$ is the propeller rotation frequency and $N_b$ is the number of blades on the propeller. 

Also, the Doppler spread of these spikes is given by
\begin{equation}
    B_D = \frac{4 \omega L \cos{\frac{\beta}{2}}\sin{\theta}}{\lambda},
    \label{equ_doppler_spread}
\end{equation}
where $L$ is the blade length, $\omega$ is the angular velocity of the propeller rotation, $\beta$ is the bistatic angle, $\theta$ is the elevation angle, and $\lambda$ is the wavelength.

\begin{figure}[b]
    \centering
    \includegraphics[width=0.45\textwidth, height=2in]{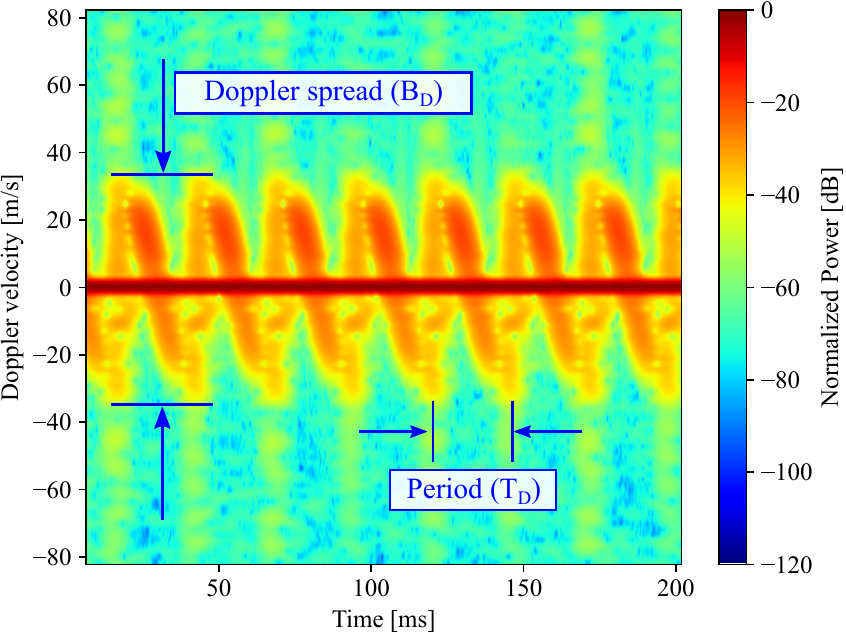}
    \caption{STFT time-frequency signature of a single VTOL rotating propeller (4096 symbols, 128 points FFT, bistatic angle = $\ang{30}$).}
    \label{fig_STFT_propeller}
\end{figure}

Although the frequency domain representation is important to help in understanding of the micro-Doppler phenomenon, a more comprehensive way of visualizing the micro-Doppler signature is the time-frequency representation. 
\autoref{fig_STFT_propeller} presents the time-frequency micro-Doppler signature of a single VTOL propeller obtained by Short-time Fourier Transform (STFT).
Here, in addition to the Doppler spread described in \eqref{equ_doppler_spread}, the periodic nature of the micro-Doppler signature in slow time can be seen, with period equal to $T_{D} = \frac{1}{N_b f_{rot}}$. 

\subsection{VTOL Micro-Doppler} \label{ssec:VTOL_mD}

\begin{figure*}[t]
\centering
\subfloat[]{%
\centering
\includegraphics[width=0.27\textwidth]{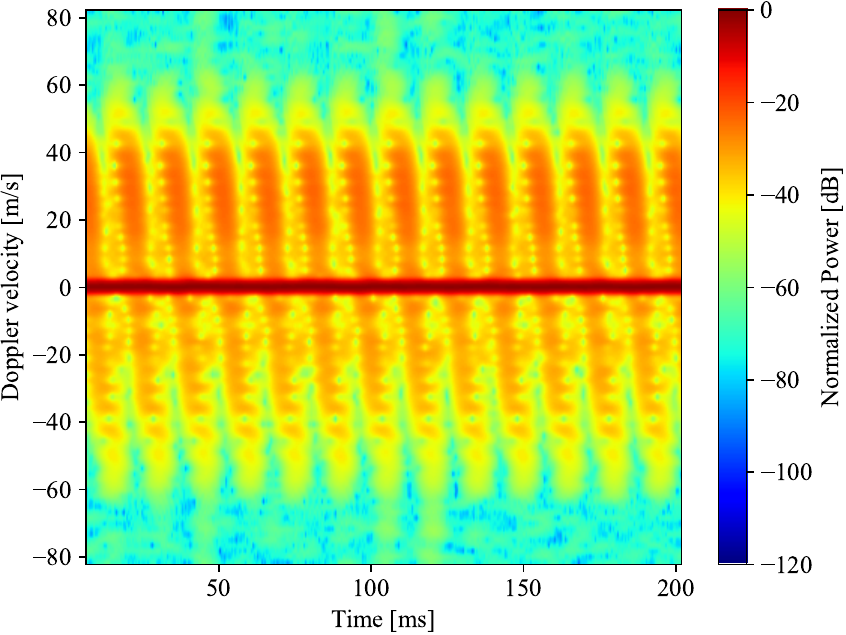}
\label{fig:6hprop_stft}
}%
~
\subfloat[]{%
\centering
\includegraphics[width=0.277\textwidth]{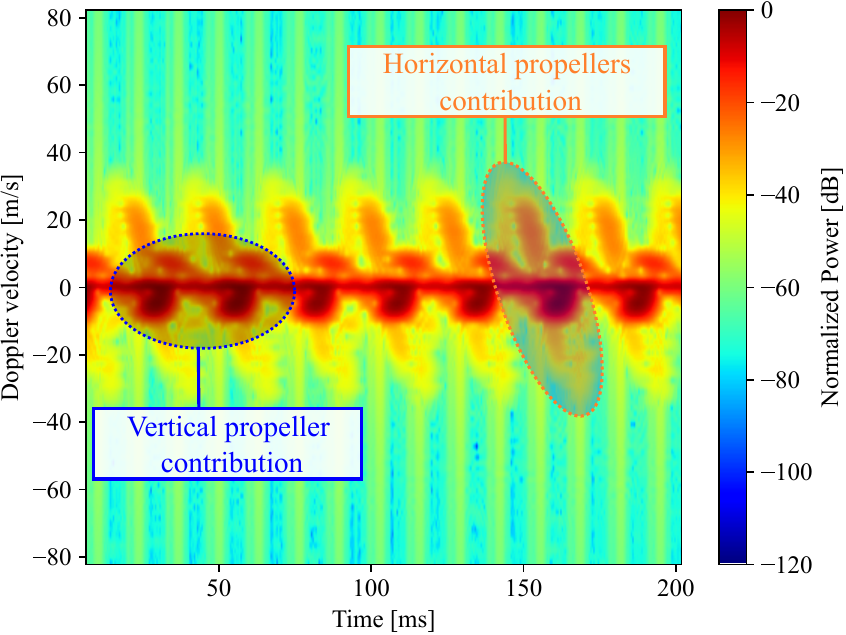}
\label{fig:7prop_stft}
}%
~
\subfloat[]{%
\centering
\includegraphics[width=0.277\textwidth]{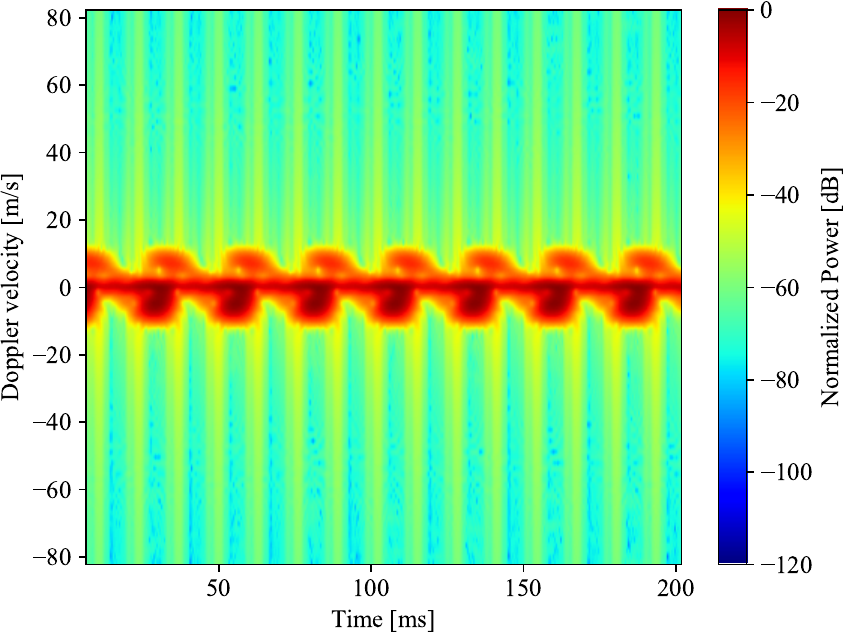}
\label{fig:1vprop_stft}
}%
\\[-2ex]
\subfloat[]{
\centering
\includegraphics[width=0.277\textwidth]{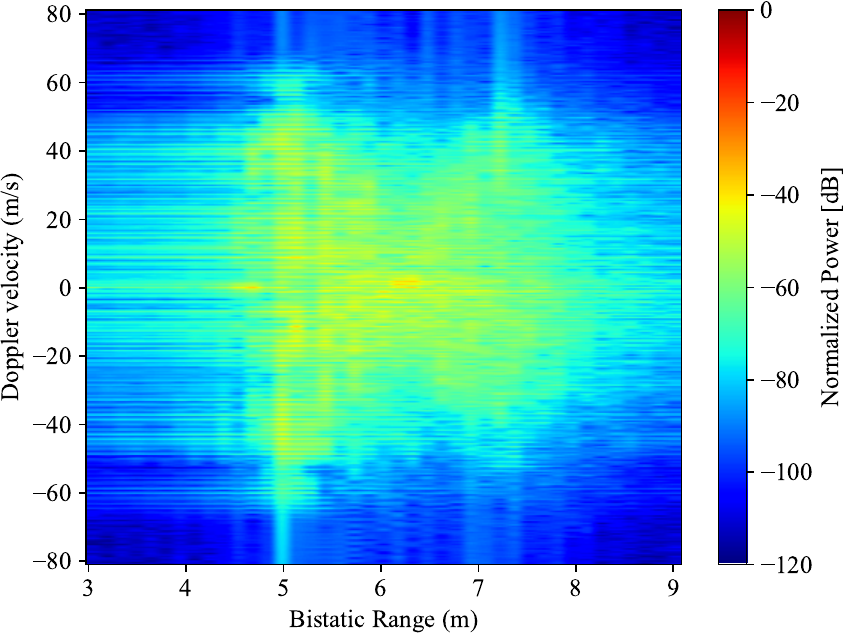}
\label{fig:6hprop_RD}
}%
~
\subfloat[]{
\centering
\includegraphics[width=0.277\textwidth]{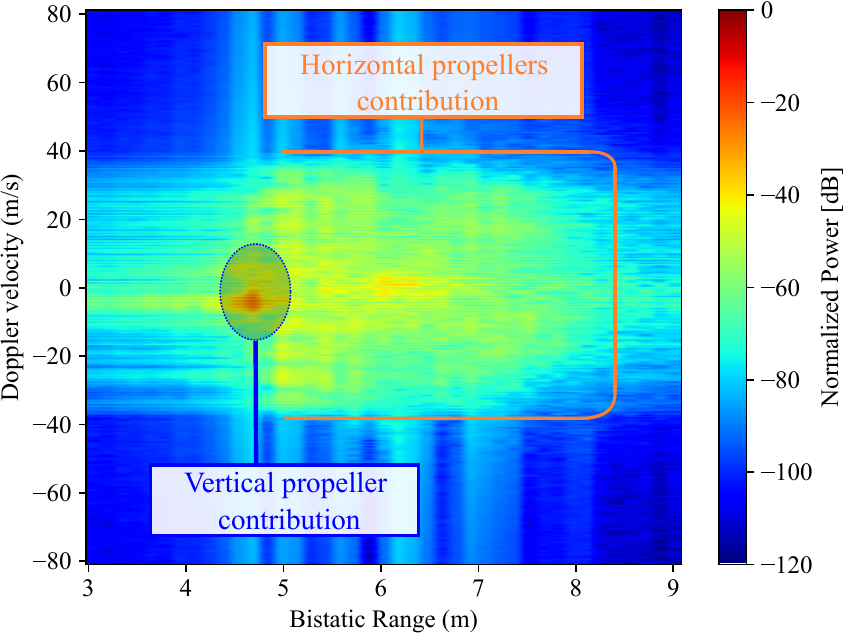}
\label{fig:7prop_RD}
}%
~
\subfloat[]{
\centering
\includegraphics[width=0.277\textwidth]{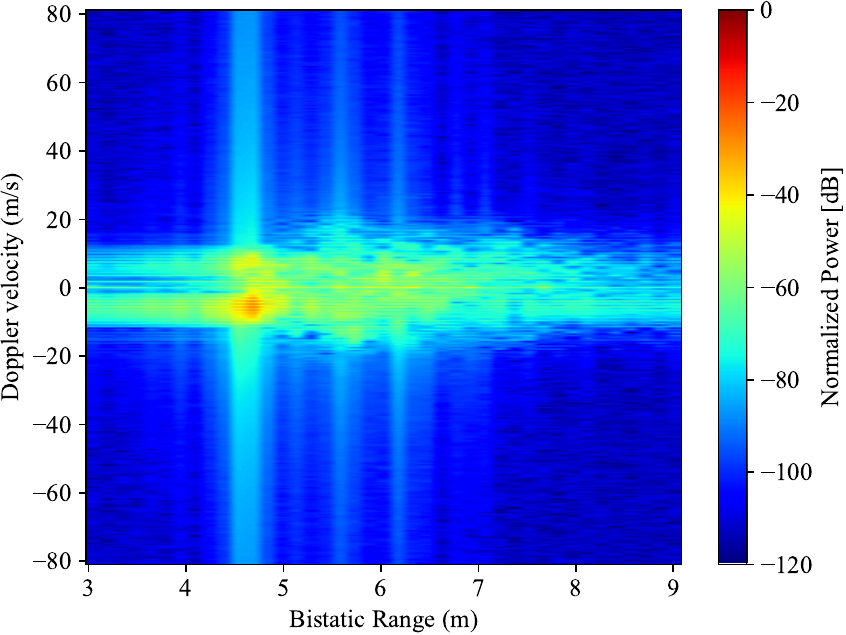}
\label{fig:1vprop_RD}
}%
\\[1.6mm]
\caption{Micro-Doppler signatures for different flight modes with $\beta = \ang{30}$:6 propellers (Take-off/landing/hovering): (a) STFT, (d) Range-Doppler map;\\
7 propellers (transition/stability): (b) STFT, (e) Range-Doppler map; 1 propeller (cruise flight): (c) STFT, (f) Range-Doppler map.}
\label{fig:VTOL_flight_modes_mD}
\vspace{-\baselineskip}
\end{figure*}

Micro-Doppler signature of VTOL is a superposition of the signatures of its propellers, each with their own Doppler spread and period in slow time.
Utilizing high signal bandwidths, the target will no longer appear in a single range bin but in multiple range bins. As a consequence, high range resolution (HRR) methods become available.
\autoref{fig:VTOL_flight_modes_mD} shows the STFT time-frequency and the HRR range-Doppler signatures of each flight mode when the bistatic angle $\beta = \ang{30}$.

\autoref{fig:6hprop_stft} and \autoref{fig:6hprop_RD} show micro-Doppler signature for six active lifting propellers and a disabled forward-thrust propeller, which refers to the  take-off, landing and hovering cases. In these figures, although it is difficult to clearly identify the contributions of each individual horizontal propeller, it is possible to identify their joint Doppler spread.

\autoref{fig:1vprop_stft} and \autoref{fig:1vprop_RD} present the signature of the forward-thrust propeller, representing the cruise flight mode. These figures show a signal that, compared to that of the six lifting propellers, has narrower Doppler spread and higher normalized power, what makes it easily distinguishable from the signature of the lifting propellers. This contrast occurs because of the difference in the geometry of this propeller, what represents a \ang{90} difference in the elevation angle $\theta$ and, consequently, results in a very different value of $B_D$ in \eqref{equ_doppler_spread}.

\autoref{fig:7prop_stft} and \ref{fig:7prop_RD} display the micro-Doppler signature when all seven propellers are present, as in the transition mode.
Again, the contributions of each individual horizontal propeller cannot be clearly identified. However, the Doppler spread of the lifting propellers, and the contribution of the vertical propeller, very similar to the one seen in \autoref{fig:1vprop_stft} and \autoref{fig:1vprop_RD}, can be easily spotted, and are clearly indicated in the figures.
The Doppler spread of the lifting propellers, between \autoref{fig:6hprop_stft} and \autoref{fig:7prop_stft} and between \autoref{fig:6hprop_RD} and \autoref{fig:7prop_RD}, are different because the rotation speeds in the measurements are also different.

In summary, as can be seen in \autoref{fig:VTOL_flight_modes_mD}, the difference in the micro-Doppler signatures for each flight mode is somewhat evident, both in time-frequency and range-Doppler representations, which makes their classification possible.

\section{Conclusions and Outlook}\label{sec:Conclusion}

This paper presents micro-Doppler signatures of VTOL propellers. 
The findings show a very distinct feature of the micro-Doppler, contributed by the different propellers.
Therefore, it can be applied to classify the flight modes of a VTOL, namely take-off, landing and hovering mode, transition mode, and cruise flight mode.

Furthermore, the rotational speed of each propeller, type of propeller, and number of propellers can also be derived from the micro-Doppler signatures.
The feature extraction and development of classifiers based on the extracted features are further works.

\section*{Acknowledgment}
The authors thank Dr.Ing. Nowack and M.Sc. Pourjafarian for their support throughout the BiRa measurement.
This research is funded by the BMBF project 6G-ICAS4Mobility (16KISK241), the Federal State of Thuringia, Germany, and by the European Social Fund (ESF) under grants 2017 FGI 0007 (project "BiRa") and 2021 FGI 0007 (project "Kreatör").

\bibliographystyle{IEEEtran}
\bibliography{main}
\end{document}